\RequirePackage{fix-cm}

\documentclass[smallextended]{svjour3}       % onecolumn (second format)
\smartqed  % flush right qed marks, e.g. at end of proof
\usepackage{xcolor}
\usepackage[nointegrals]{wasysym}
\usepackage[T1]{fontenc}
\usepackage{graphicx}
\usepackage{aas-macros}
\usepackage{upgreek}
\usepackage{amsmath,amssymb}%
\usepackage{listings}
\usepackage{subcaption}
\usepackage{longtable}
\usepackage[pagewise]{lineno} % For line numbers
\newcommand{\RAPOC}{\texttt{RAPOC}}

\begin{document}

\title{\RAPOC\ : the Rosseland and Planck opacity converter %\thanks{Grants or other notes
%about the article that should go on the front page should be
%placed here. General acknowledgments should be placed at the end of the article.}
}
\subtitle{A user-friendly and fast opacity program for Python}

\titlerunning{\RAPOC}        % if too long for running head

\author{Lorenzo V. Mugnai$^{1,2}$ \and 
    Darius Modirrousta-Galian$^{3,2,4}$}

\authorrunning{Lorenzo V. Mugnai \and Darius Modirrousta-Galian}

\institute{
\email{lorenzo.mugnai@uniroma1.it} \\
$^{1}$Dipartimento di Fisica, La Sapienza Universit\`a di Roma, Piazzale Aldo Moro 2, 00185,Roma, Italy \\
$^{2}$ INAF – Osservatorio Astronomico di Palermo, Piazza del Parlamento 1, I-90134 Palermo, Italy \\
$^{3}$Department of Earth and Planetary Sciences, Yale University
New Haven, CT 06511, USA \\
$^{4}$ University of Palermo, Department of Physics and Chemistry, Via Archirafi 36, Palermo, Italy}

\date{Received: date / Accepted: date}
% The correct dates will be entered by the editor

\maketitle

\begin{abstract}
\RAPOC\ (Rosseland and Planck Opacity Converter) is a Python 3 code that calculates Rosseland and Planck mean opacities (RPMs) from wavelength-dependent opacities for a given temperature, pressure, and wavelength range. In addition to being user-friendly and rapid, \RAPOC\ can interpolate between discrete data points, making it flexible and widely applicable to the astrophysical and Earth-sciences fields, as well as in engineering. For the input data, \RAPOC\ can use \textit{ExoMol} and \textit{DACE} data, or any user-defined data, provided that it is in a readable format. In this paper, we present the \RAPOC\ code and compare its calculated Rosseland and Planck mean opacities with other values found in the literature. The \RAPOC\ code is open-source and available on Pypi and GitHub.

\keywords{Opacity \and Exoplanets \and Atmospheres \and Infrared \and Visible}
% \PACS{PACS code1 \and PACS code2 \and more}
% \subclass{MSC code1 \and MSC code2 \and more}
\end{abstract}

\section{Introduction}
\label{intro}
Our understanding of extra-solar planetary systems has grown significantly since the first exoplanet was discovered in 1995 \cite{Mayor1995}. One major aspect of this field, is the analysis, exploration, and modelling of planetary atmospheres; all of which require a careful treatment of opacities. For example, atmospheric spectroscopy makes use of wavelength-dependent opacities to determine the chemical constituents present in the observable part of the atmosphere of exoplanets. Projects such as \textit{ExoMol}\footnote{https://www.exomol.com} \cite{Tennyson2016}, \textit{HITRAN}\footnote{https://hitran.org/} \cite{Gordon2017}, \textit{HITEMP}\footnote{https://hitran.org/hitemp/} \cite{Rothman2010}, which are dedicated to generating line-lists for spectroscopy, have facilitated our aim of understanding the atmospheric properties of other worlds \cite{Tsiaras_2018,Tsiaras2019,Edwards2020,Skaf2020,Pluriel2020,Guilluy2021,Mugnai2021,Giacobbe2021,Changeat_2021}. In contrast, when theoretically modelling an exoplanetary atmosphere, opacities are often used to estimate the global temperature profile and the location of the radiative-convective boundary. This approach can work with wavelength-dependent or wavelength-averaged opacities, in which the former is generally accepted to be a more rigorous and accurate representation of real systems than the latter.
Although the aforementioned projects primarily focus on line-lists, they can be converted into opacity tables \cite{Yurchenko2018} that can be more straightforward to operate with; \textit{ExoMol} \cite{Chubb2021}  and \textit{DACE}\footnote{https://dace.unige.ch/opacityDatabase/} (Data and Analysis Center for Exoplanets) \cite{Grimm2021} provide such conversions. Whereas wavelength-dependent opacities can be used in theoretical models, they require computationally-intensive simulations \cite{Pluriel2020,Fortney2007,Nettelmann2011,Petralia2020}; it is, therefore, common to rely on Rosseland and Planck mean opacities (shortened to RM for the former, PM for the latter, and RPM when referring to both) that only depend on the temperature and pressure of the system. The use of RPMs has several benefits that include (1) their wavelength independence makes them simpler and faster to use, (2) they can be implemented in Grey and semi-Grey models to provide a reasonable estimation of the temperature structure of astrophysical and planetary environments, and (3) such modelling can provide exact solutions.

Whereas we do not explore Grey and semi-Grey approaches in this paper, it is pertinent to discuss them because they are widely used in various academic fields, and they make use of RPMs. Grey and semi-Grey models are approximate analytical solutions to the radiative transfer analyses of gaseous environments, which are defined as using either one (the infrared) or two (the infrared and visible) wavelength-averaged opacities, respectively. In atmospheric sciences, such approaches were popularised by Sir Arthur Eddington \cite{Eddington1916} and then expanded upon by various others \cite{Chandrasekhar1935,King1955,King1956,Chandrasekhar1960,Matsui1986,Weaver1995,Pujol2003,Hubeny2003,Chevallier2007,Hansen2008,Burrows2010,Guillot2010,Shaviv2011}. A rigorous comparison of the Grey and semi-Grey models available in the literature has previously been done \cite{Parmentier2014}, so it will not be explored in this study. With the recent launch of the \textit{JWST} \cite{Greene2016} on December 24th 2021, and several upcoming astronomical missions like \textit{Ariel} \cite{Tinetti2018,Tinetti2021} and \textit{Twinkle} \cite{Edwards2019}, there is a strong motivation for further exploring extra-solar planetary atmospheres. We also recognise that grey and semi-grey approaches may also be used in planetary formation models \cite{Pollack1985,Henning1995,Henning1996} and engineering \cite{Viskanta1987,Moreno1991,Wang2014}.

Despite the advantages of Grey and semi-Grey approaches, RPM data are not commonly produced. This has motivated some researchers to assume constant values \cite{Guillot2010}, or adopt simple analytic approximations \cite{Kurosaki2014}. In light of this problem, we present \RAPOC, a Python program that converts the readily available wavelength-dependent opacities into RPMs.

\textit{Caveat -- }the \RAPOC\ code is not a replacement for more rigorous radiative transfer approaches\cite{Fortney2007,Nettelmann2011,Petralia2020,Pluriel2020}. Instead, it is built to provide pressure and temperature-dependent RPMs in a spectral range of choice so that Grey and semi-Grey models can include more complex opacity behaviour. This may increase the efficacy of such approaches, assuming that the pressure-temperature models used are appropriate approximations of reality. Whenever and wherever possible, the authors recommend using more rigorous approaches instead of Grey and semi-Grey approximations. 

\section{What is \RAPOC ?}
\label{sec:1}

\RAPOC\ (Rosseland and Planck Opacity converter) is a fast and user-friendly program that is fully written in Python 3 and converts wavelength-dependent opacities into RPMs as a function of the temperature and pressure for the wavelength range of choice. RPMs are usually given as a function of density and temperature (not pressure) \cite{Semenov2003,Mayer2005}, because opacities are defined as $k_{\nu} \equiv \kappa_{\nu} \rho \equiv \alpha_{\nu} n$, where $k_{\nu}$ is the volume opacity (not further referenced in this study), $\kappa_{\nu}$ is the mass opacity, $\alpha_{\nu}$ is  the extinction coefficient, $\rho$ is the density, and $n$ is the number density. \textit{ExoMol} and \textit{DACE} provide opacities as a function of temperature, pressure and wavelength, but not density;this is because their opacities are computed from line lists that are pressure-broadened. More information on the input data can be found in sect.~\ref{sec:inputs}. \RAPOC\ is publicly available on Pypi\footnote{https://pypi.org/project/rapoc/}, so it can be installed using the \textit{pip} command
\begin{lstlisting}[language=bash]
$ pip install rapoc
\end{lstlisting}
or it can be compiled directly from the source, and downloaded from the GitHub repository\footnote{https://github.com/ExObsSim/Rapoc-public} with
\begin{lstlisting}[language=bash]
$ cd Rapoc
$ pip install .
\end{lstlisting}
All data generated in this paper used \RAPOC\ version 1.0.5. This version, and all future versions, are available on the GitHub repository. For the complete and extensive \RAPOC\ guide, please refer to the software documentation\footnote{https://rapoc-public.readthedocs.io/en/latest/}.

\subsection{Rosseland mean opacity}

The Rosseland mean opacity (RM) is defined as \cite{Lenzuni1991}
\begin{equation} \label{eq:rosseland}
\frac{1}{\kappa_r} = \frac{\int_{0}^{\infty} \kappa_{\nu}^{-1} u(\nu, T) d\nu} {\int_{0}^{\infty} u(\nu, T) d\nu},
\end{equation}
where $\kappa_\nu$ is the opacity provided by the input data at a given frequency $\nu$, and $u(\nu, T)$ is the Planck black body derivative with respect to the temperature $T$. Because opacity data is generally not available across the entire electromagnetic spectrum, a shorter range is selected (i.e., multigroup opacities). With \RAPOC, the user selects the frequency range of interest $(\nu_1, \nu_2)$ to compute the mean opacity. Eq.~\ref{eq:rosseland} can therefore be rewritten as
\begin{equation}\label{eq:rapoc_rosseland}
\frac{1}{\kappa_r} \simeq \frac{\int_{\nu_1}^{\nu_2} \kappa_{\nu}^{-1} u(\nu, T) d\nu} {\int_{\nu_1}^{\nu_2} u(\nu, T) d\nu}.
\end{equation}

Due to the definition of RM, if the wavelength-dependent opacity, $\kappa_{v}$, were zero at a given wavelength, Eq.~\ref{eq:rosseland} would be numerically undefined. This causes an error, so we included a fail-safe correction where the zero is replaced by an arbitrarily small value. This \textit{ad-hoc} correction keeps the code functional.

\subsection{Planck mean opacity}
The Planck mean opacity (PM) is defined as \cite{Lenzuni1991}
\begin{equation} \label{eq:planck}
\kappa_p = \frac{\int_{0}^{\infty} \kappa_{\nu} B_\nu(T) d\nu}{\int_{0}^{\infty} B_\nu(T) d\nu},
\end{equation}
where $B_\nu(T)$ is the Planck black body law computed at temperature $T$. For the same reasons given previously, Eq.~\ref{eq:planck} is rewritten as
\begin{equation}\label{eq:rapoc_planck}
\kappa_p \simeq \frac{\int_{\nu_1}^{\nu_2} \kappa_{\nu} B_\nu(T) d\nu}{\int_{\nu_1}^{\nu_2} B_\nu(T) d\nu}.
\end{equation}

\subsection{Inputs}
\label{sec:inputs}

The first step in the code is to load the opacity data to initialise the \texttt{Model} class. This data can be provided as a file or in a custom-made Python dictionary format. The data must contain an opacity table with a corresponding list of pressures, temperatures and wavenumbers (or wavelengths or frequencies) so that the opacities can be sampled. 

\subsubsection{Input data file}

As of the writing of this paper, the \RAPOC\ code only accepts \textit{ExoMol} cross-sections in the TauREx format \cite{Al-Refaie2020} format, and \textit{DACE} opacities \cite{Grimm2021} as input data\footnote{Future versions of \RAPOC\ will implement a load function for different input files by using the dedicated \texttt{FileLoader} class.}. 

\paragraph{\textit{ExoMol} cross-section (TauRex format).}
Raw opacity data is available for a large sample of molecules on the \textit{ExoMol} website\footnote{https://exomol.com/data/data-types/opacity/} \cite{Barber2014,Yurchenko2017,Polyansky2018,Coles2019,Chubb2021}. \textit{ExoMol} data is structured as a grid of pressures, temperatures, wavenumbers, and cross-sections. By using the \texttt{units} module of the \texttt{Astropy} package \cite{astropy}, \RAPOC\ attaches units to the data so that conversions are straightforward. By default, pressure is expressed in $Pa$ and temperature in $K$; the wavenumber grid (expressed in $1/cm$) is converted into wavelengths ($\mu m$), and frequencies ($Hz$). All previously mentioned information is stored in the \texttt{Model} class. Finally, the cross-sections contained in the \textit{ExoMol} file are loaded. These are given in units of $cm^2/molecule$, so to convert them into opacities, the mass of the molecule is retrieved using the \texttt{molmass} Python package\footnote{https://pypi.org/project/molmass/} and then divide through by the absorption table. After taking into consideration the necessary unit conversions, the opacities expressed in $m^2/kg$ are obtained. The steps mentioned above lead to a three-dimensional Numpy array \cite{oliphant_numpy} table of opacities with indexes corresponding to the pressure, temperature, and wavenumber grids, respectively. These are stored in the \texttt{Model} class under the attribute \texttt{opacities}.

\paragraph{\textit{DACE} opacities.} 

The \textit{DACE} database collects line-lists produced by projects like \textit{ExoMol}, \textit{HITRAN}, and \textit{HITEMP}, and converts them into opacities using the \texttt{HELIOS-K} opacity calculator \cite{Grimm2015}. The opacity data can be downloaded from the \textit{DACE} database in a directory for each molecule containing the binary files. Each of these files contains the opacity as a function of the wavenumber, pressure, and temperature. The downloaded opacities come in units of $cm^2/g$. \RAPOC\ accepts the directory address as input and parses the contained files to build a three-dimensional Numpy array of opacities ordered for pressure, temperature, and wavenumber. All units are then converted into SI units and, subsequently, all of the aforementioned information is stored in the \texttt{Model} class.

\subsubsection{Input Python dictionary}

As previously mentioned, instead of an input file, the user may use a Python dictionary as the required input. The dictionary content must be of the same type as the one described for the input file due to \RAPOC\ handling the contained data in the same way. Therefore, the dictionary must contain the following five entries: (1) \texttt{mol} -- a string for the molecule name, (2) \texttt{pressure} -- an array for the pressure grid data, (3) \texttt{temperature} -- an array for the temperature grid data, (4) \texttt{wavenumber} -- an array for the wavenumber grid data, and (5) \texttt{opacities} -- a three-dimensional array of the opacities (in units of area over mass) ordered by pressures, temperatures and wavenumbers. Optionally, the dictionary can contain the molecular mass, under (6) \texttt{mol\_mass} key. If this key is not present, \RAPOC\ will compute this quantity automatically. We point the reader to Table~\ref{tab:input_dictionary} for a schematic representation of the dictionary structure or, alternatively, to the \RAPOC\ documentation for a full description.

\begin{table}[]
    \centering
    \begin{tabular}{p{0.15\linewidth} | p{0.18\linewidth} | p{0.4\linewidth}|p{0.10\linewidth}}
    \hline
        \textbf{keyword} & \textbf{data type} & \textbf{description} &\textbf{required} \\ 
    \hline
        \texttt{mol} & \textit{string} & Molecule name. & Yes\\
        \texttt{pressure} & \textit{numpy.array or Quantity} & Pressure grid.  & Yes \\ 
        \texttt{temperature} & \textit{numpy.array} or \textit{Quantity} & Temperature grid.  & Yes\\ 
        \texttt{wavenumber} & \textit{numpy.array} or \textit{Quantity}  & Wavenumber grid.  & Yes\\ 
        \texttt{opacities} &\textit{numpy.array} or \textit{Quantity}  & Three-dimensional array of the opacities ordered by pressure (axis 0), temperature (axis 1), and wavenumbers (axis 2).  & Yes \\ 
        \texttt{mol\_mass} & \textit{float} or \textit{Quantity} & Molecular weight.  & No\\
    \hline
    \end{tabular}
    \caption{Input dictionary structure. A custom made Python dictionary can be used as a \texttt{RAPOC} input if it contains the indicated keywords.}
    \label{tab:input_dictionary}
\end{table}

\subsection{Rayleigh scattering} \label{sec:rayleigh}
\RAPOC\ includes a module for producing Rayleigh scattering RPMs for the atomic species given in table~\ref{tab:polarisabilities} that is found in Appendix \ref{sec:rayleigh_tab}. The Rayleigh scattering wavelength-dependent opacity is given by \cite{CRC92,Modirrousta2021},
\begin{equation}
    k_{Ray}(\lambda) = \frac{128 \pi^5 }{3 \mu \lambda^4} \cdot \alpha^2,
\end{equation}
where $\mu$ is the atomic mass expressed in SI units and $\alpha$ is the static average electric dipole polarisability of the gaseous species being considered, which is given in Table~\ref{tab:polarisabilities}. The resulting $k_{Ray}(\lambda)$ values are expressed in $m^2/kg$ and processed in the RPM modules to compute their respective mean opacities. Although the Rayleigh scattering opacity is independent of temperature, RPMs are not due to the black-body equation (or its derivative), as shown in Eq. \ref{eq:rosseland} and \ref{eq:planck}. Conversely, pressure is not required, so whether or not a pressure is inserted, \RAPOC\ will ignore it.

\subsection{Estimation algorithms}

The \RAPOC\ code offers two estimation methods. For the first method, given the requested pressure $P$ and temperature $T$ input by the user, \RAPOC\ finds the closest pressure and temperature in the data grid, extracts the opacity data, and computes the desired mean opacity in the frequency range (or wavelength or wavenumber) of choice (i.e. $\nu_1, \nu_2$). Eq. \ref{eq:rapoc_rosseland} and Eq. \ref{eq:rapoc_planck} are used to calculate the RMs or PMs respectively.

The second method consists of an interpolation of the estimated RPM values. For this method, \RAPOC\ first produces a map by computing the model mean opacity (RM or PM with Eq.~\ref{eq:rosseland} or Eq.~\ref{eq:planck}, respectively) at the indicated frequency band $(\nu_1, \nu_2)$ for every pressure and temperature available in the data. To make the code faster, once the map is built, the code will not reproduce it, unless the user changes the investigated frequency bands in a successive iteration. This map can be used to interpolate the model opacity values for given pressures and temperatures as long as they are within the bounds of the data grids. The aim of this code is to compute RPMs from given input data in a reliable and efficient manner. Therefore, extrapolation methods are not implemented into \RAPOC. Hence, inserting an input pressure or temperature outside the data range will result in an error. Nevertheless, the interpolation is handled by the \texttt{Scipy} \texttt{griddata} module \cite{scipy} using the \textit{linear} mode described in the documentation. As input data usually contains a wide range of pressures, \RAPOC\ also contains a \textit{loglinear} mode, which is the same as \textit{linear} but the pressure is in a logarithmic form when the interpolation is made. We stress that the Scipy \texttt{griddata} algorithms should be used carefully as the quality of the interpolation is dependent on the local environment as well as the mode requested, so unphysical estimations may occur. Both estimation algorithms allow the user to compute the RPMs for either a single input pressure and temperature, or for a grid, by giving a list of pressures and temperatures as inputs.

\subsection{Outputs}

\begin{figure}[th]
\centering
	\begin{subfigure}[b]{0.48\textwidth}
		\centering
		\includegraphics[width=\textwidth]{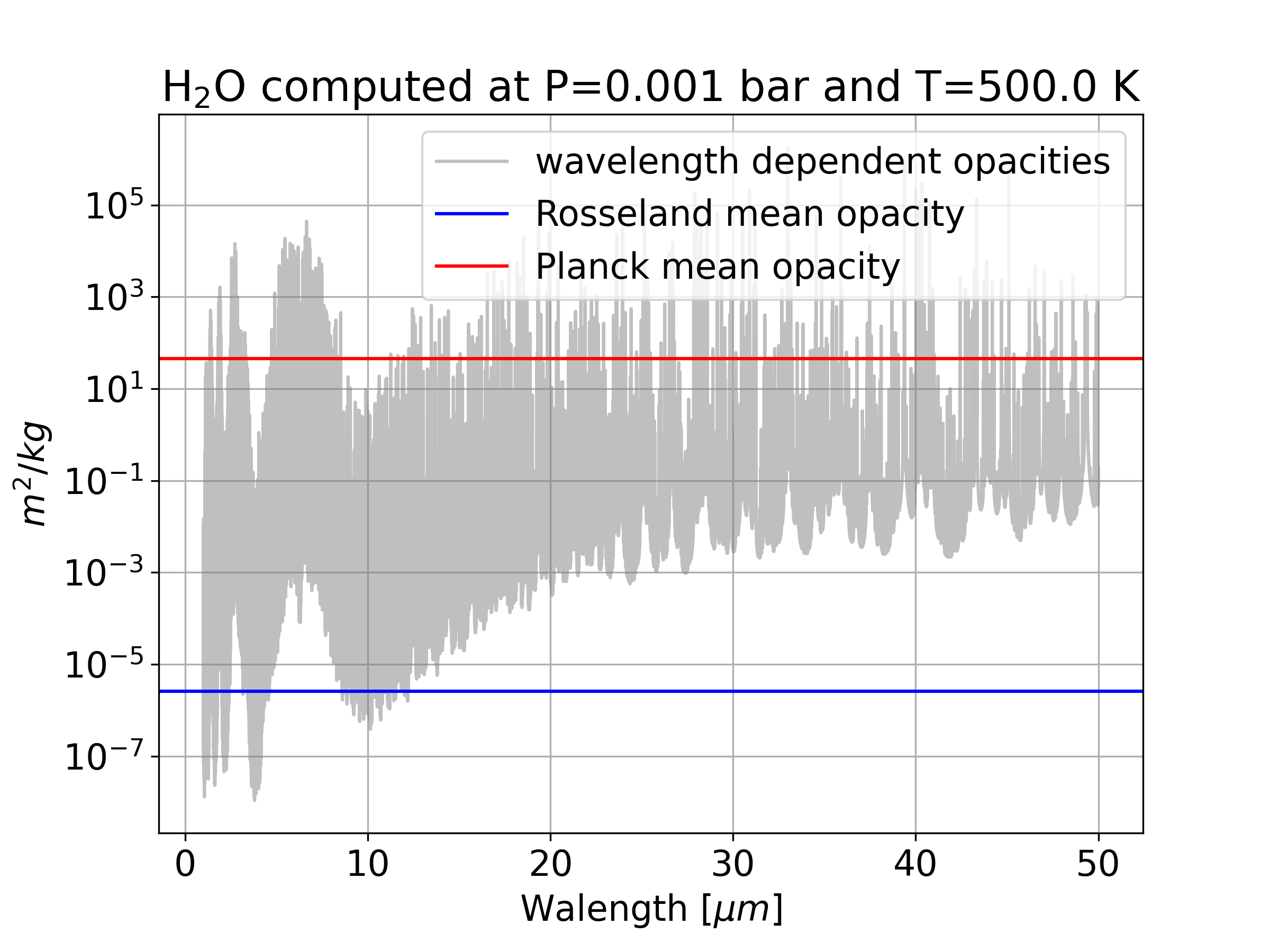}
		\caption{H$_2$O mean opacities in the $\rm 1 - 50 \, \mu m $ range at $\rm P=0.001 \, bar$ and $\rm T=500 \,K$.}
	\end{subfigure}
	\hfill
	\begin{subfigure}[b]{0.48\textwidth}
		\centering
		\includegraphics[width=\textwidth]{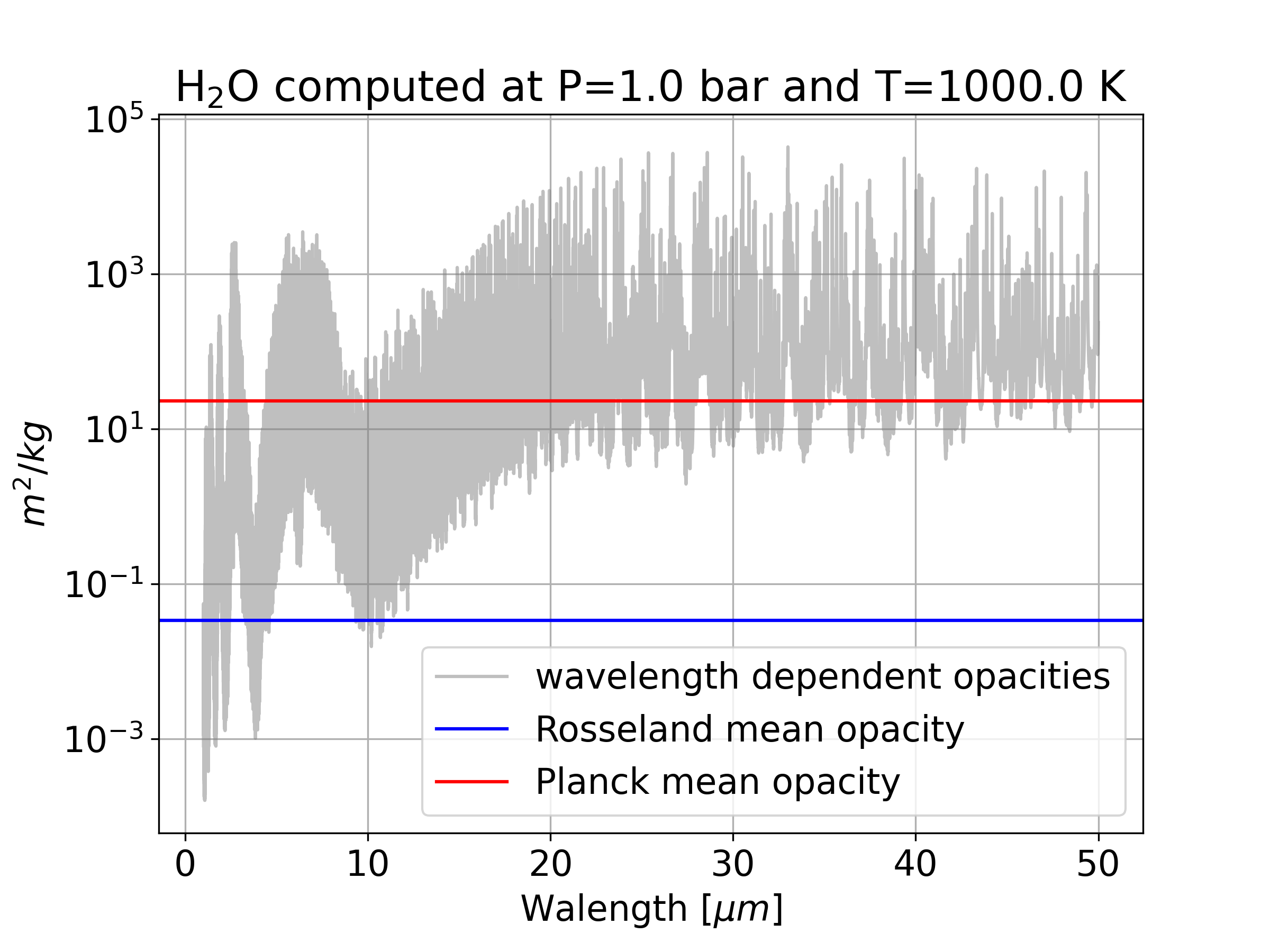}
		\caption{H$_2$O mean opacities in the $\rm 1 - 50 \, \mu m $ range at $\rm P=1 \, bar$ and $\rm T=1000 \,K$.}	
	\end{subfigure}

	\begin{subfigure}[b]{0.48\textwidth}
		\centering
		\includegraphics[width=\textwidth]{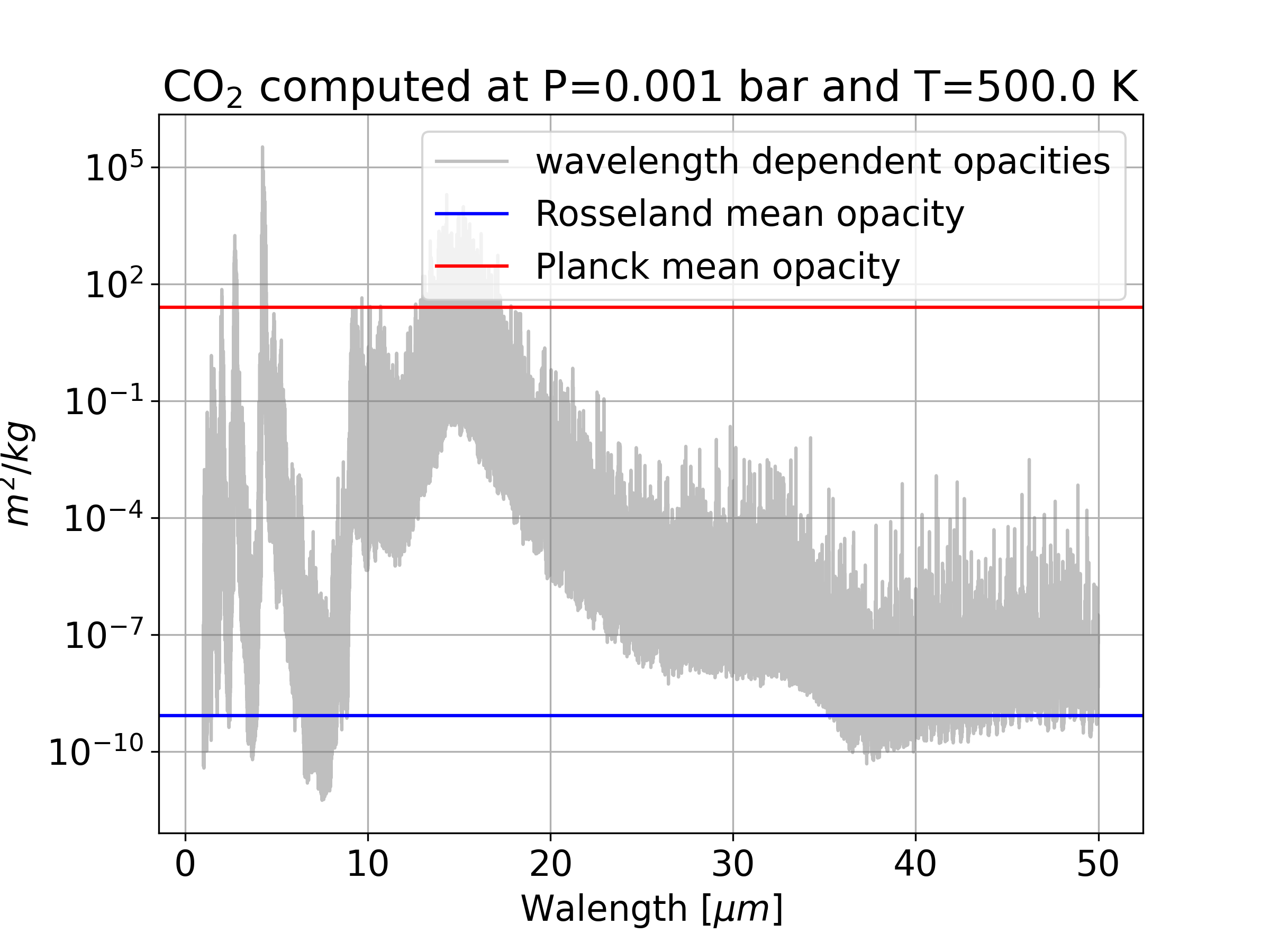}
		\caption{CO$_2$ mean opacities in the $\rm 1 - 50 \, \mu m $ range at $\rm P=0.001 \, bar$ and $\rm T=500 \,K$.}	
	\end{subfigure}
	\hfill
	\begin{subfigure}[b]{0.48\textwidth}
		\centering
		\includegraphics[width=\textwidth]{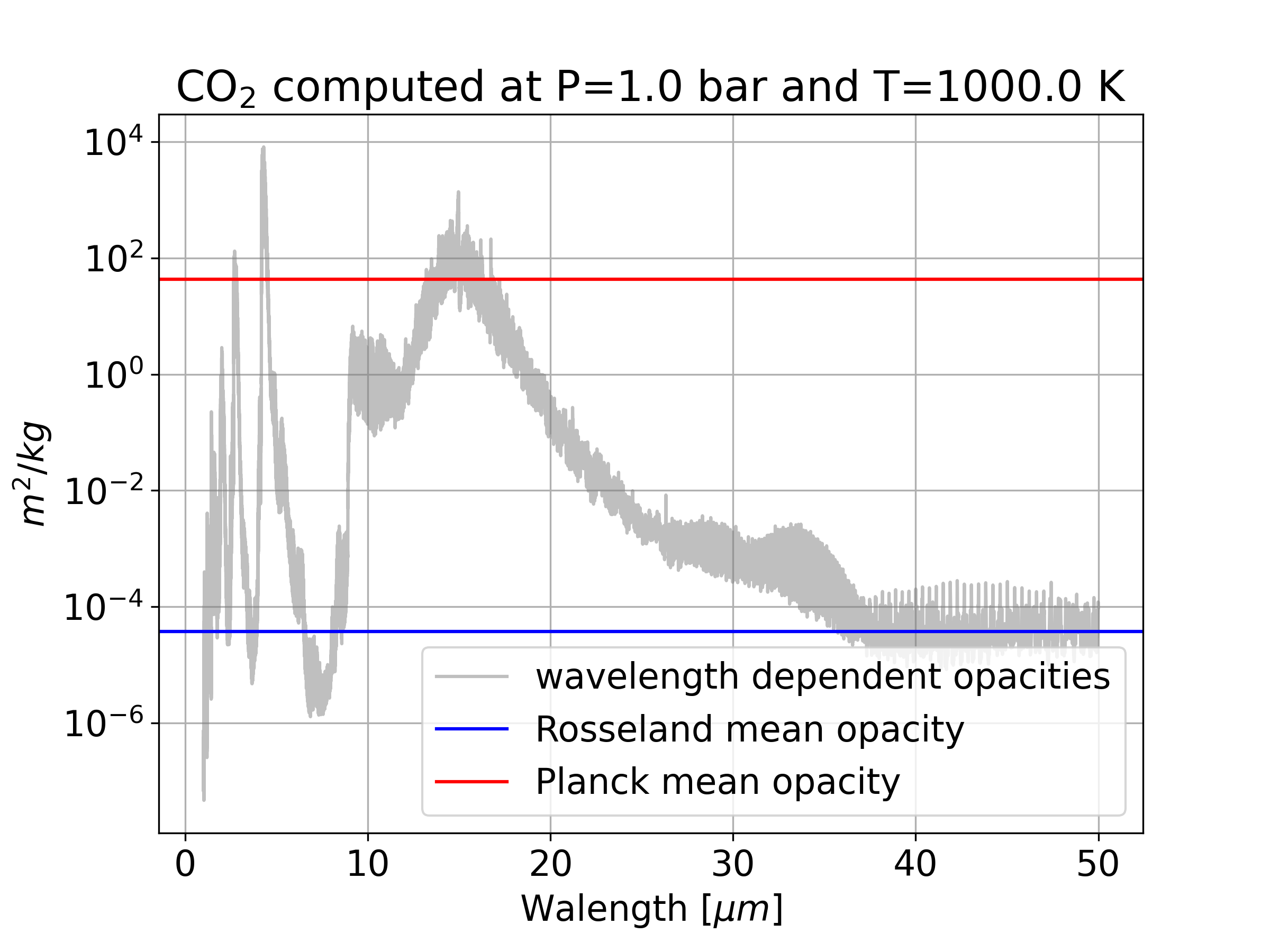}
		\caption{CO$_2$ mean opacities in the $\rm 1 - 50 \, \mu m $ range at $\rm P=1 \, bar$ and $\rm T=1000 \,K$.}
	\end{subfigure}

	\caption{The mean opacities computed by \RAPOC\ for four different cases. In each panel the grey line represents the input data opacities (\textit{ExoMol}) with their corresponding pressures and temperatures in the given wavelength range. The blue and red lines are the computed RMs and PMs, respectively. These have been estimated with the \textit{closest} method. In the top row, the opacities of water are shown, while the bottom row is for methane. The right column reports the results for $\rm P=0.001 \, bar$ and $\rm T=500 \, K$, while the left row shows the equivalent for $\rm P=1 \, bar$ and $\rm T=1000 \, K$. In all of the panels the wavelength range is $\rm 1 - 50 \, \mu m $.}
    \label{fig:rapoc_example}
\end{figure}

\begin{figure}
\centering
	\begin{subfigure}[b]{\textwidth}
		\centering
		\includegraphics[width=\textwidth]{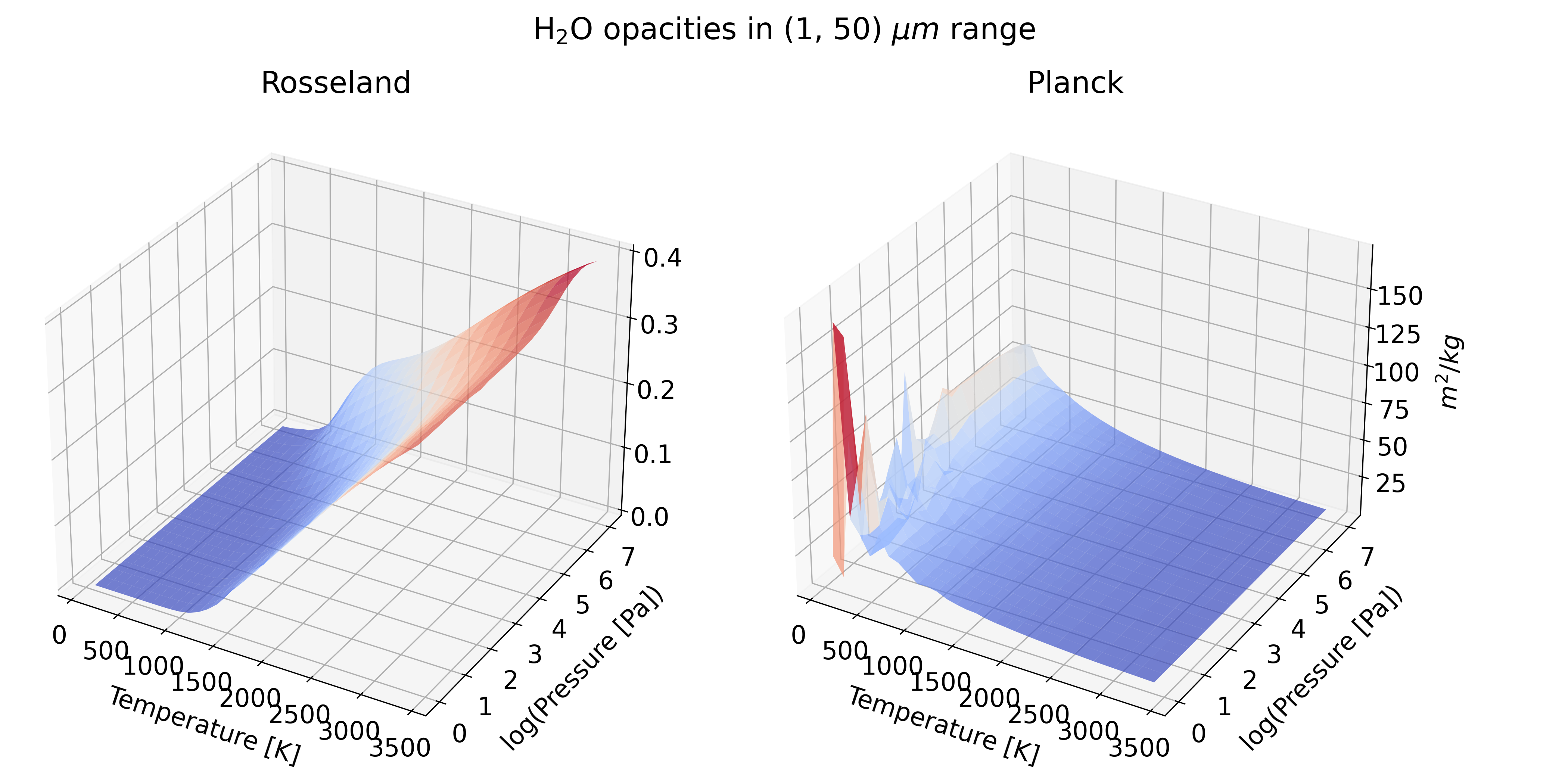}
		\caption{H$_2$O mean opacities in the $\rm 1 - 50 \, \mu m $ range.}
	\end{subfigure}

	\begin{subfigure}[b]{\textwidth}
		\centering
		\includegraphics[width=\textwidth]{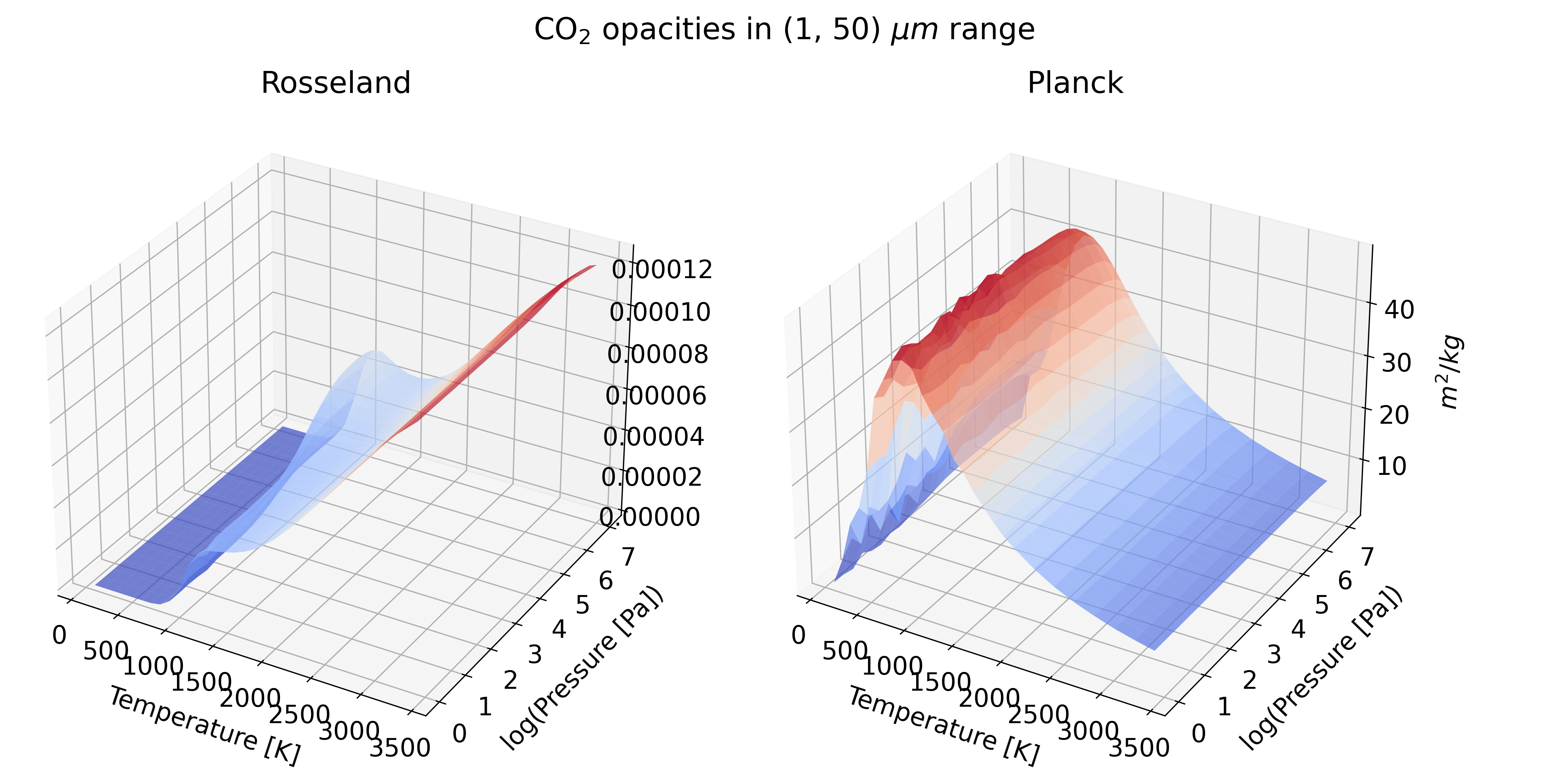}
		\caption{CO$_2$ mean opacities in the $\rm 1 - 50 \, \mu m $ range.}	
	\end{subfigure}

	\caption{Opacity map produced with \RAPOC\ for H$_2$O\cite{Polyansky2018} (top row) and CO$_2$\cite{Yurchenko2020} (bottom row) over the $\rm 1-50 \, \mu m$ range using \textit{ExoMol} input data. RMs are reported on the left column, while PMs are reported on the right column.}
    \label{fig:rapoc_map_example}
\end{figure}

\begin{figure}
    \centering
    \includegraphics[width=\textwidth]{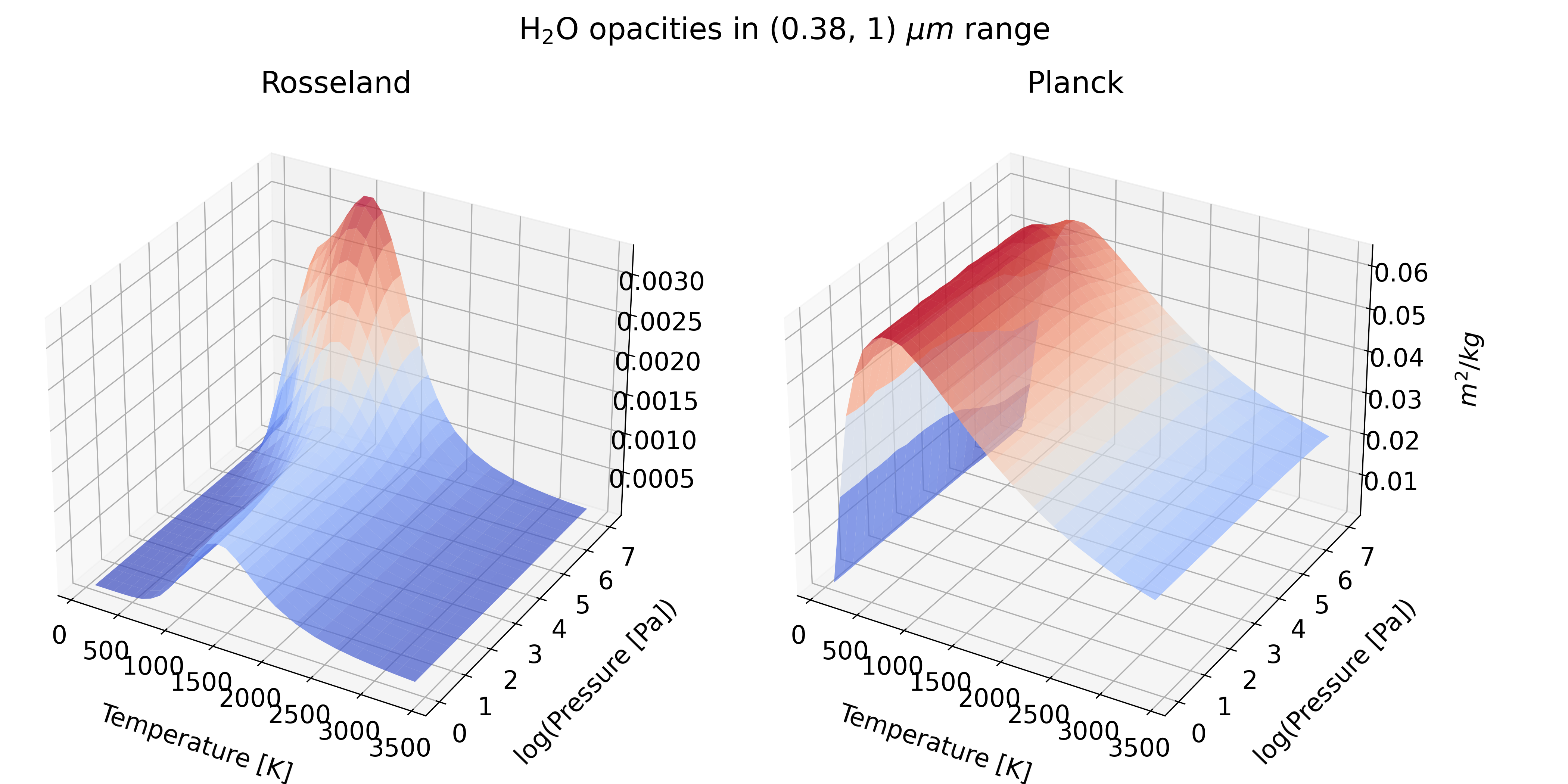}

	\caption{Opacities map produced with \RAPOC\ for H$_2$O\cite{Polyansky2018} over the $\rm 0.38-1 \, \mu m$ range from \textit{ExoMol} input data. RMs are reported on the left, while PMs are reported on the right.}
    \label{fig:rapoc_map_example_0307}
\end{figure}

An example of the possible outputs is shown in Table~\ref{tab:rapoc_estimates}. In the table, estimates are provided for \textit{ExoMol}'s water \cite{Polyansky2018} and carbon dioxide \cite{Yurchenko2020} data using the \textit{linear} method. In Fig.~\ref{fig:rapoc_example}, we show the RPM values estimated by \RAPOC\ compared to the input opacity as a function of wavelength, temperature and pressure. Using \RAPOC, one can generate a map of RMs and PMs for each combination of pressure and temperature available in the input data; examples of these maps are shown in Fig.~\ref{fig:rapoc_map_example} and \ref{fig:rapoc_map_example_0307}. The two figures demonstrate that opacities usually do not have a monotonic behaviour in their respective pressure and temperature grids. By comparing Fig.~\ref{fig:rapoc_map_example} with Fig.~\ref{fig:rapoc_map_example_0307} one sees that the estimated opacities are strongly dependent on the wavelength range considered. 

\begin{table}[]
    \centering
    \begin{tabular}{c|c|c|c|c}
       \hline\noalign{\smallskip}
		{\bf \centering Molecule} & {\bf \centering P} & {\bf \centering T} & {\bf \centering RM} & {\bf \centering PM} \\ 
		& {\centering [bar]} & { \centering [K]} & {\centering [$\mathrm{m}^2/\mathrm{kg}$]} & {\centering [$\mathrm{m}^2/\mathrm{kg}$]} \\ 
       \hline\noalign{\smallskip}

$\rm H_2O$ & $0.01$ & $1000$ & $0.0043$ & $22.8322$ \\ 
$\rm H_2O$ & $1.0$ & $500$ & $0.0012$ & $40.4905$ \\ 
$\rm H_2O$ & $1.0$ & $1000$ & $0.0341$ & $23.0859$ \\ 
$\rm H_2O$ & $1.0$ & $1500$ & $0.1002$ & $14.7822$ \\ 
$\rm H_2O$ & $1.0$ & $2500$ & $0.2484$ & $8.1571$ \\ 
$\rm H_2O$ & $100.0$ & $1000$ & $0.1185$ & $22.3517$ \\ 
$\rm CO_2$ & $0.01$ & $1000$ & $0.0000$ & $43.3707$ \\ 
$\rm CO_2$ & $1.0$ & $500$ & $0.0000$ & $29.2340$ \\ 
$\rm CO_2$ & $1.0$ & $1000$ & $0.0000$ & $43.6445$ \\ 
$\rm CO_2$ & $1.0$ & $1500$ & $0.0000$ & $28.1226$ \\ 
$\rm CO_2$ & $1.0$ & $2500$ & $0.0001$ & $12.1506$ \\ 
$\rm CO_2$ & $100.0$ & $1000$ & $0.0000$ & $39.5779$ \\  
      	\noalign{\smallskip}\hline

    \end{tabular}
    \caption{RPMs estimated by \RAPOC\ for different temperatures and pressures in the $\rm 1-50 \, \mu m$ wavelength range for water and methane using \textit{ExoMol} data. This estimation has been performed using the \textit{linear} method.}
    \label{tab:rapoc_estimates}
\end{table}

\section{Discussion}

% In sect.~\ref{sec:realistic} and \ref{sec:otherRPM} we present the realisations of \RAPOC\ in the context of more rigorous approaches and other RPM data in the literature. 
\subsection{Other Opacity Rosseland \& Planck Opacity Sources} \label{sec:comp_literature}

There are various resources for RPMs in the literature, but most focus on primordial gas mixtures with different metallicities \cite{Cox1976,Alexander1989,Lenzuni1991,Alexander1994,Iglesias1996,Mayer2005,Freedman2008,Freedman2014}. Whereas the values of mixtures are useful for modelling planetary formation or stellar interiors, they are not as applicable to planetary atmospheres. We, therefore, focus on papers providing RPMs for individual molecules because they allow for a straightforward comparison with the RPM values provided by \RAPOC. Most of the individual molecule RPM values present in the literature are estimated directly from line-lists \cite{Badescu2010,Kurosaki2014}. Alternatively, \RAPOC\ uses precomputed opacities for single molecules to estimate their wavelength-averaged values, which allows for faster and easier computations, and a straightforward integration into other codes. \RAPOC, therefore, relies on precomputed data, such as the one provided by \textit{ExoMol} and \textit{DACE}, instead of line lists. Furthermore, if the opacities of a gas mixture are required, the user must manually account for the contributions of the individual species calculate by \RAPOC. In the following, we compare the estimates obtained by \RAPOC\ with others found in the literature.   

\begin{figure}[ht]
    \centering
    \includegraphics[width = \columnwidth]{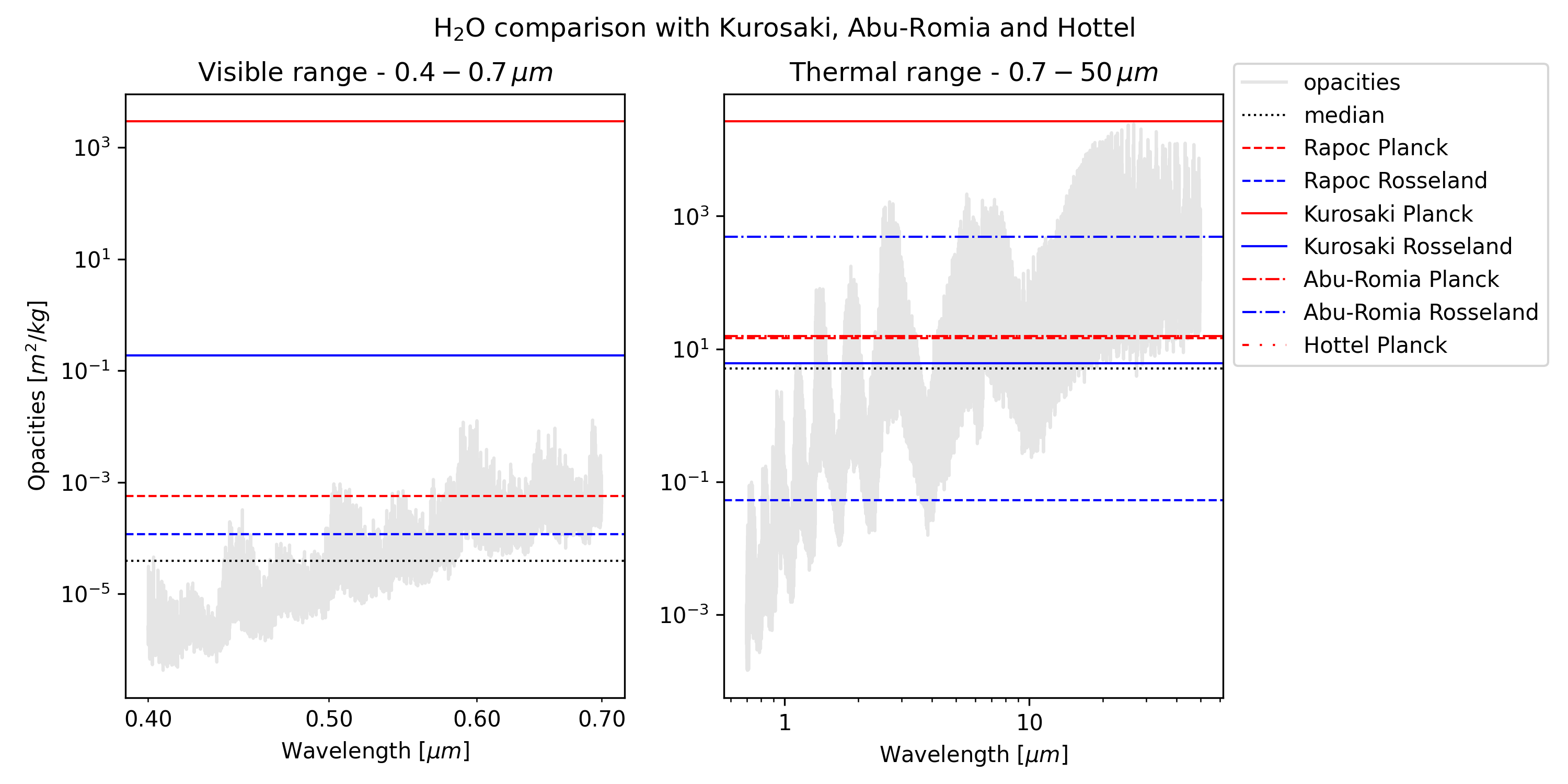}
    \caption{Comparison between Kurosaki et al. \cite{Kurosaki2014}, Abu-Romia \& Tien \cite{aburomia1967}, Hottel \cite{Hottel1954} and \RAPOC. The shaded lines in both plots represent the raw data loaded from \textit{ExoMol}'s water opacities \cite{Polyansky2018}. The blue lines are Rosseland Mean Opacities with the filled lines being from Kurosaki et al., the dash-dotted line from Abu-Romia \& Tien, and the dashed lines from \RAPOC. The red lines are Planck Mean Opacities with the filled lines being from Kurosaki et al., the dash-dotted line from Abu-Romia \& Tien, the dash-dot-dotted line from Hottel, and the dashed lines from \RAPOC. The black dotted line is the median value of the raw wavelength dependent opacities. The left panel is for the visible wavelength range ($0.3$ to $0.7 \, \mu m$) and right panel is for the IR wavelength range ($0.7$ to $50 \, \mu m$). Both panels use the same pressure (1.01325 bar) and temperature (1500 K).}
    \label{fig:comp_kurosaki}
\end{figure}

We compare our RPM opacity estimations for water vapor with those of Hottel \cite{Hottel1954}, Abu-Romia \& Tien \cite{aburomia1967}, and Kurosaki et al. \cite{Kurosaki2014}. Hottel estimated the IR Planck mean opacities from emissivity data, whereas Abu-Romia \& Tien found IR RPMs from spectral data using selected
bands in the $2.7 - 20 \, \mu m$ range, which contribute appreciably to the emitted energy. Kurosaki et al., however, produces a monotonic power-law fit (their Eqs.~A.5--8) for estimating water RPMs in the visible and thermal wavelengths using HITRAN data. The power-law approximation presented in Kurosaki et al. has been tuned for two wavelength ranges: visible ($0.4 - 0.7 \, \mu m$) and thermal ($0.7 - 100 \,\mu m$). For a comparison with \RAPOC, we estimate RPMs with $0.4 - 0.7 \, \mu m$ and $0.7 - 50 \, \mu m$ wavelength ranges for visible and thermal range respectively. Our comparison is found in Fig.~\ref{fig:comp_kurosaki}. Because Abu-Romia \& Tien and Hottel only provide results the IR range, we only show Kurosaki et al. for the visible range.
We are aware that for simple molecules such as $\rm H_{2}$ that are weakly absorbing in the infrared and visible wavelengths, the collisional absorption may be crudely approximately by a power law as a function of pressure and temperature. However, as soon as a hydrogen gas is slightly enriched by other molecules, the power-law approximation begins to fail \cite{Freedman2008,Freedman2014}. In addition, for molecules like $\rm H_{2}O$ and $\rm CO_{2}$, there are other sources of opacity such as electronic transitions, molecular rotations, and vibrations, meaning that the opacity is not at all monotonic. Because of this, and the different wavelength ranges considered, the model by Kurosaki et al. predicts opacities that differ by up to five orders of magnitude from what is estimated by \RAPOC. Fig.~\ref{fig:comp_kurosaki} shows how Kurosaki et al. predicts opacities that are significantly greater than the wavelength-dependent values available from \textit{ExoMol}.

As previously mentioned, Abu-Romia \& Tien and Hottel results are only applicable to the IR range. The data reported in Abu-Romia \& Tien \cite{aburomia1967} are displayed in figures with temperature on the x-axis (in units of Rankine) and opacity (as inverse feet) in the y-axis. We convert their estimates by dividing their opacities by the local gas density, which is estimated using the ideal gas equation
\begin{equation}
    \rho = \frac{M P}{R T},
\end{equation}
where $M$ is the molar mass of the gas ($M_{H_2O} = 18.01528 \, g/mol$ for water), $P$ is the pressure, $R$ is the ideal gas constant, and $T$ is the temperature.
Fig. \ref{fig:comp_kurosaki} shows that \RAPOC's Planck Mean Opacity estimate is compatible with the value reported in Abu-Romia \& Tien and Hottel. The Rosseland Mean Opacity given by Abu-Romia \& Tien is, however, several orders of magnitudes larger than both the value estimated by Kurosaki et al. and \RAPOC. 

\begin{figure}[ht]
    \centering
    \includegraphics[width = \columnwidth]{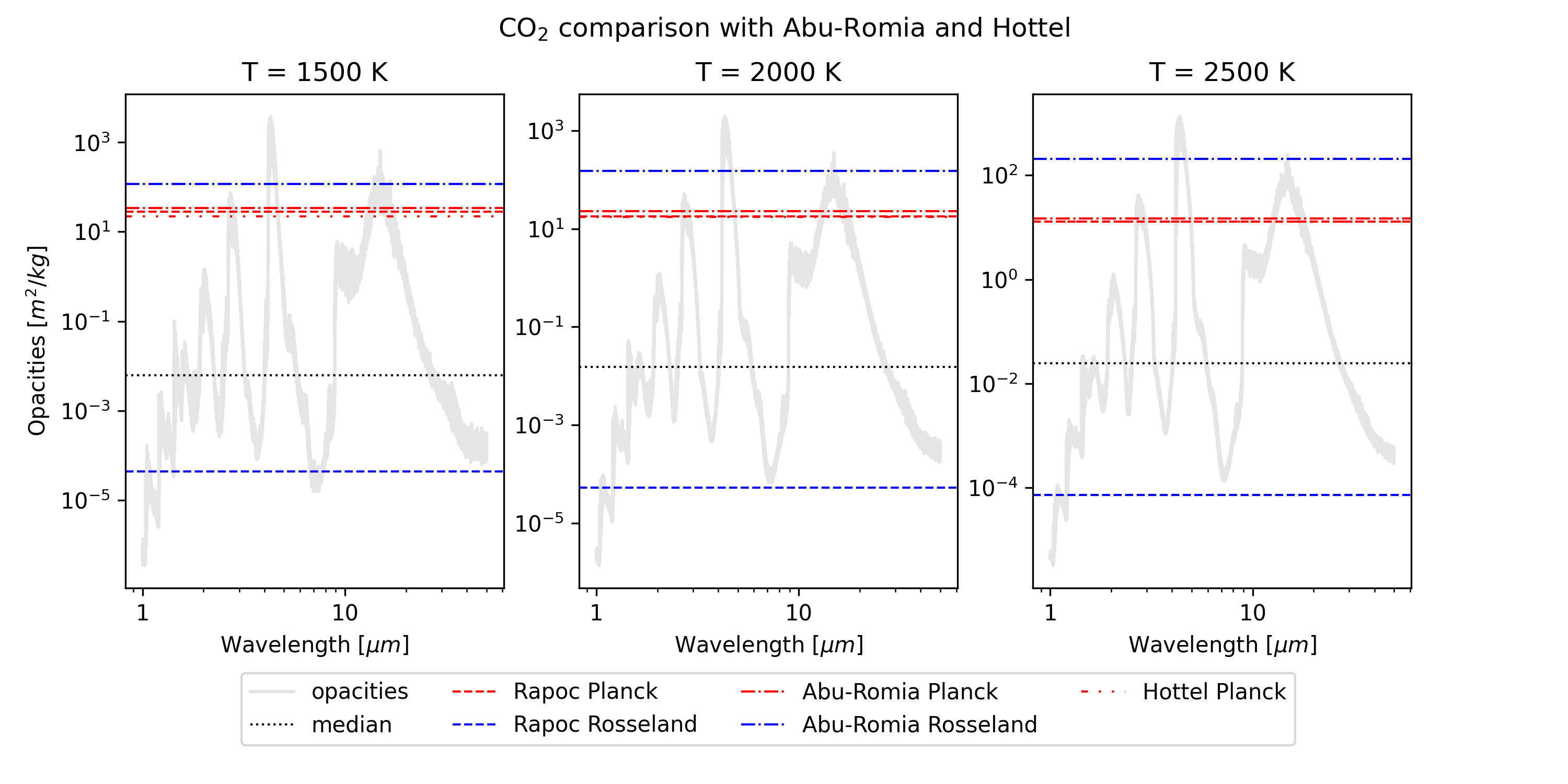}
    \caption{Comparison between Abu-Romia \& Tien \cite{aburomia1967}, Hottel \cite{Hottel1954} and \RAPOC. The shaded lines in all plots represent the raw data loaded from \textit{ExoMol}'s carbon dioxide opacities \cite{Yurchenko2020}. The blue lines are Rosseland Mean Opacities with the dash-dotted lines being from Abu-Romia \& Tien, and the dashed lines from \RAPOC\ . The red lines are Planck Mean Opacities with  the dash-dot-dotted lines being from Hottel, the dash-dotted lines from Abu-Romia \& Tien, and the dashed being from \RAPOC. The black dotted line is the median value of the raw wavelength dependent opacities. The three panels refer to different gas temperatures: right is for $T=1500 \, K$, centre is for $T=2000 \, K$, and right is for $T=2500 \, K$. All panels use the same pressure (1.01325 bar).}
    \label{fig:comp_abu}
\end{figure}

Regarding $\rm CO_{2}$, we compare the Planck opacities calculated by \RAPOC\ with those given in Abu-Romia \& Tien \cite{aburomia1967} and Hottel \cite{Hottel1954}; the Rosseland opacities are compared to those of Badescu \cite{Badescu2010}. The comparison for the PM is shown in Fig.~\ref{fig:comp_abu}, where the Planck opacities are given for three different temperatures. The opacities from Abu-Romia \& Tien \cite{aburomia1967} were extracted from their graphs, as described previously, but by using the molar mass of carbon dioxide $M_{CO_2} = 44.01 \, g/mol$. As shown in Fig~\ref{fig:comp_abu}, the \RAPOC\ Planck opacities are consistent with those of Abu-Romia \& Tien and Hottel.

\begin{figure}[ht]
    \centering
    \includegraphics[width = \columnwidth]{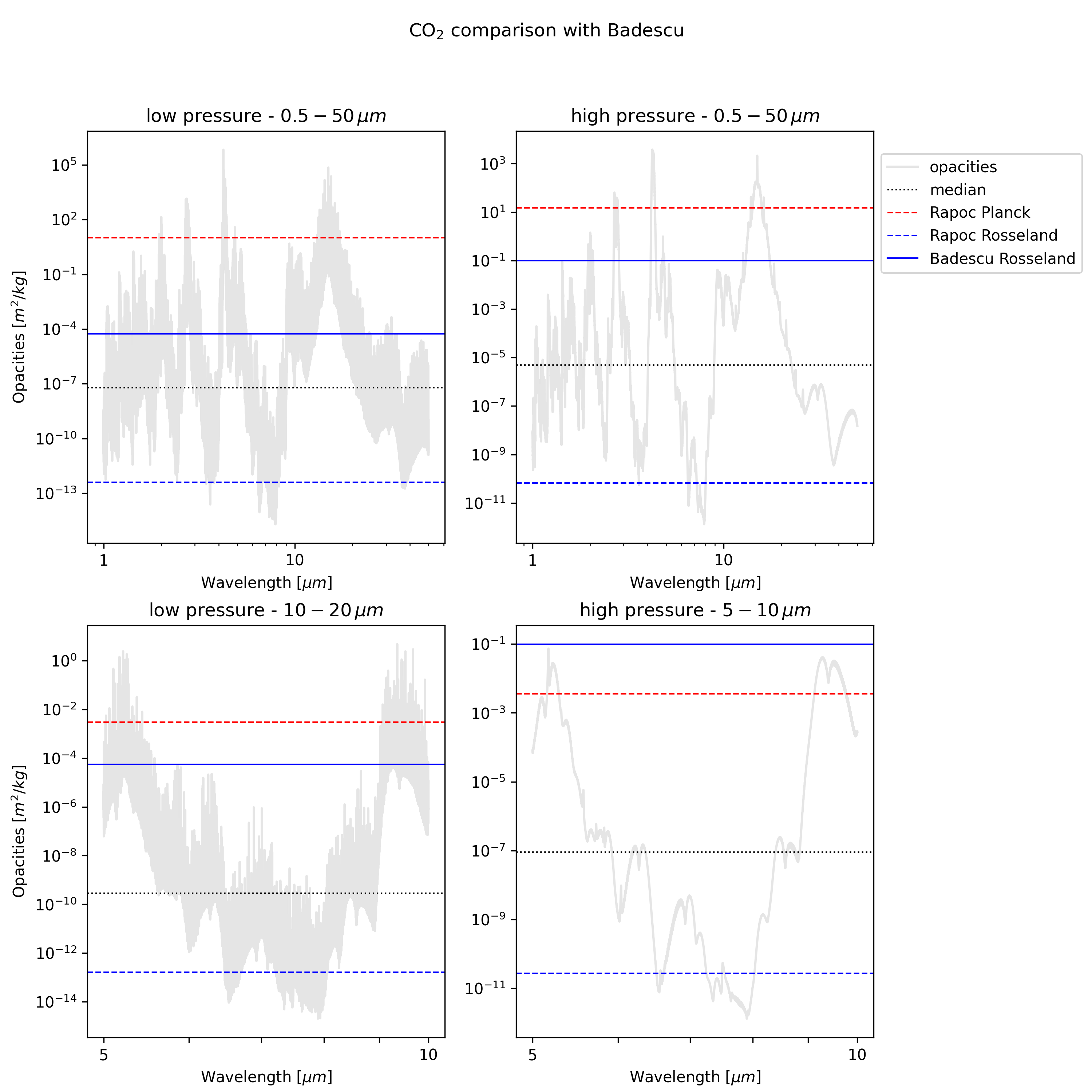}
    \caption{Comparison between Badescu \cite{Badescu2010} and \RAPOC\ estimates. The shaded lines in all plots represent the raw data loaded from \textit{ExoMol}'s carbon dioxide opacities \cite{Yurchenko2020}. The blue lines are Rosseland Mean Opacities with the filled lines from Badescu and the dashed lines from \RAPOC. The red dashed lines are \RAPOC's Planck Mean opacities, and the black dotted lines are the median value of the raw wavelength dependent opacities. The left column is for low pressure ($567 \cdot 10^{-3}$ bar) and the right column is for high pressure (11.467 bar); both columns use the same temperature (300 K). The wavelength range is different in the two rows as the top row uses $0.5$ to $50 \, \mu m$ range, while the bottom row uses $5$ to $10 \, \mu m$.}
    \label{fig:comp_baduscu_co2}
\end{figure}

For the Rosseland mean opacities, Table~6 of Badescu \cite{Badescu2010} is considered. In their calculation, a wavelength range of $\rm 0.5~\mu m$ to $\rm 100~\mu m$ was used, which is beyond the limit provided by \textit{ExoMol} data \cite{Yurchenko2020}. Hence, a wavelength range of $0.5~\mu m$ to $\rm 50~\mu m$ will be adopted when making the comparison. The results are shown in the first row of Fig.~\ref{fig:comp_baduscu_co2}. The figure shows that Badescu's estimates are closer to the median value of the wavelength dependent opacities from \textit{ExoMol} than what \RAPOC\ calculates. The bottom row of the same figure reports the same estimates performed on the $5-10~\mu m$ wavelength range. A major advantage of the \RAPOC\ code is that the wavelength range can be specified, whereas using Badescu's values are given for a set wavelength range.

For the water and carbon dioxide cases, there are significant differences between the RPMs given by \RAPOC\ and those available in the literature; the only exception being the Planck mean opacities that are consistent with those of Abu-Romia \& Tien \cite{aburomia1967} and Hottel \cite{Hottel1954}. These differences are the result of different wavelength ranges investigated, or the adoption of simple analytic approximations, such as the power-law fit introduced in Kurosaki et al. \cite{Kurosaki2014}. The major advantage of \RAPOC\ is that it provides a flexible and systematic avenue for calculating RPMs with widely available input data. As shown in the above, this flexibility is coupled with \RAPOC's ability to better represent the weighted mean opacity of a gaseous species at a given spectral window, or across a large range than the other approaches in the literature. Despite these advantages, \RAPOC is dependent on the input data (excluding the Rayleigh scattering opacities), and it cannot extrapolate outside the given wavelength, temperature, and pressure bounds.

\subsection{Limitations with Rosseland and Planck Mean Opacities}

Whereas RPMs have their uses, they are also limited. For instance, in optically-thin environments, RPMs may overestimate the opacities present as photons could traverse through `spectral windows', which might be very different from a few strong opacity regions. Furthermore, RM and PM have different functional forms corresponding to the different averages they are providing. The RM opacity uses the derivative of the Planck distribution as the weighting function, which it then uses to find the harmonic mean of the opacity. Consequently, RMs are extremely sensitive to the opacity minima and can provide erroneous values if a molecule is fully transparent at a given wavelength. Conversely, PM opacity uses the Planck function as the weighting function and then finds the arithmetic mean, so it is strongly affected by the more opaque regions of the spectrum. Due to their different averaging prescriptions, RM and PM opacities can differ by over two orders of magnitude which, depending on thermodynamic properties of the system, could lead to substantially different temperature profiles.

\section{Summary and Conclusion}

In this paper we present the \RAPOC\ code that is able to convert wavelength-dependent opacity data into Rosseland and Planck mean opacities (RPMs) in an efficient manner. Our code is fully written in Python and publicly available on GitHub and Pypi. \RAPOC\ uses \textit{ExoMol} and \textit{DACE} data, but user-defined data can also be used as an input as long as it is within a readable format. By incorporating the pressure and temperature dependence of RPMs, \RAPOC\ provides a more complex treatment of the mean opacities than what is sometimes used in the literature, notably, assuming constant values or adopting simple analytic formulations. Whereas RPMs should not be used as a replacement for more rigorous opacity analyses, they have certain benefits. For example, RPMs allow one to use Grey or semi-Grey models when analysing gaseous environments, which are simpler and have exact solutions. We note that \RAPOC\ should not be used as an alternative to more thorough approaches such as those using wavelength-dependent opacities. However, for simpler models, \RAPOC\ provides a prescription for evaluating wavelength-dependent opacities, which can be used for exploring a larger parameter space, as well as benchmark testing.

\section{Applications}

\appendix

\section{Polarisabilities Used for the Rayleigh Scattering Opacities} \label{sec:rayleigh_tab}

\begin{longtable}{|c|c|c|c|c|c|}

\hline
\textbf{Atom} & \textbf{Polarisability} & \textbf{Atom} & \textbf{Polarisability} & \textbf{Atom} & \textbf{Polarisability} \\

& {\bf \centering [$\rm 10^{-30}~m^{3}$]} & & {\bf \centering [$\rm 10^{-30}~m^{3}$]} & & {\bf \centering [$\rm 10^{-30}~m^{3}$]} \\
\hline\noalign{\smallskip}
\endfirsthead
H  & 0.666793 & Br & 3.05   & Tm & 21.8  \\
He & 0.204956 & Kr & 2.4844 & Yb & 21    \\
Li & 24.3     & Rb & 47.3   & Lu & 21.9  \\
Be & 5.6      & Sr & 27.6   & Hf & 16.2  \\
B  & 3.03     & Y  & 22.7   & Ta & 13.1  \\
C  & 1.76     & Zr & 17.9   & W  & 11.1  \\
N  & 1.1      & Nb & 15.7   & Re & 9.7   \\
O  & 0.802    & Mo & 12.8   & Os & 8.5   \\
F  & 0.557    & Tc & 11.4   & Ir & 7.6   \\
Ne & 0.3956   & Ru & 9.6    & Pt & 6.5   \\
Na & 24.11    & Rh & 8.6    & Au & 5.8   \\
Mg & 10.6     & Pb & 4.8    & Hg & 5.02  \\
Al & 6.8      & Ag & 7.2    & Tl & 7.6   \\
Si & 5.38     & Cd & 7.36   & Pb & 6.8   \\
P  & 3.63     & In & 10.2   & Bi & 7.4   \\
S  & 2.9      & Sn & 7.7    & Po & 6.8   \\
Cl & 2.18     & Sb & 6.6    & At & 6     \\
Ar & 1.6411   & Te & 5.5    & Rn & 5.3   \\
K  & 43.4     & I  & 5.35   & Fr & 47.1  \\
Ca & 22.8     & Xe & 4.044  & Ra & 38.3  \\
Sc & 17.8     & Cs & 59.42  & Ac & 32.1  \\
Ti & 14.6     & Ba & 39.7   & Th & 32.1  \\
V  & 12.4     & La & 31.1   & Pa & 25.4  \\
Cr & 11.6     & Ce & 29.6   & U  & 24.9  \\
Mn & 9.4      & Pr & 28.2   & Np & 24.8  \\
Fe & 8.4      & Nd & 31.4   & Pu & 24.5  \\
Co & 7.5      & Pm & 30.1   & Am & 23.3  \\
Ni & 6.8      & Sm & 28.8   & Cm & 23    \\
Cu & 6.2      & Eu & 27.7   & Bk & 22.7  \\
Zn & 5.75     & Gd & 23.5   & Cf & 20.5  \\
Ga & 8.12     & Tb & 25.5   & Es & 19.7  \\
Ge & 6.07     & Dy & 24.5   & Fm & 23.8  \\
As & 4.31     & Ho & 23.6   & Md & 18.2  \\
Se & 3.77     & Er & 22.7   & No & 17.5  \\
\hline\noalign{\smallskip}
\caption{The static average electric dipole polarisabilities for ground state atoms used in \RAPOC. An analysis of the values with their associated references can be found in the referenced CRC handbook\cite{CRC92}.} % needs to go inside 
\label{tab:polarisabilities}
\end{longtable}

\begin{acknowledgements}
The authors thank the anonymous referee for their thorough evaluating. We also thank G. Micela and E. Pascale for useful comments. This work has been supported by the ARIEL ASI-INAF agreement n. 2018.22.HH.0 and n. 2021.5.HH.0.
\end{acknowledgements}

% Authors must disclose all relationships or interests that 
% could have direct or potential influence or impart bias on 
% the work: 
%
\section*{Conflict of interest}

The authors declare that they have no conflict of interest.

\section*{Data Availability Statement}
The molecular opacities data-sets analysed during the current study are available in the \textit{ExoMol} repository (https://www.exomol.com).
The RPM data generated in this work can be reproduced by using \RAPOC\ code, version 1.0.5, which is publicly available on GitHub (https://github.com/ExObsSim/Rapoc-public) and Pyppi (https://pypi.org/project/rapoc/).

% BibTeX users please use one of
%\bibliographystyle{spbasic}      % basic style, author-year citations
%\bibliographystyle{spmpsci}      % mathematics and physical sciences
%\bibliographystyle{spphys}       % APS-like style for physics
%\bibliography{}   % name your BibTeX data base

\bibliographystyle{spphys}
\bibliography{mainv3}
%
% and use \bibitem to create references. Consult the Instructions
% for authors for reference list style.

\end{document}